\documentclass[iop]{emulateapj}
\usepackage{amssymb}
\usepackage{mathrsfs}
\usepackage{rotating}
\usepackage{graphicx}

\newcommand{\fn}[1]{\footnote{\scriptsize{#1}}} 

\newcommand{\Fig}[1]{Fig{#1}.}  

\begin{document} 

\journalinfo{\textit{Astrophys.~J.~Lett.} \textbf{718}, L92--L96 (2010).}
\submitted{}
\title{Physical characteristics and non-keplerian orbital motion\\of ``propeller'' moons embedded in Saturn's rings} 

\author{Matthew~S.~Tiscareno$^1$, Joseph~A.~Burns$^{1,2}$, Miodrag~Srem\v{c}evi\'c$^3$, Kevin~Beurle$^4$, Matthew~M.~Hedman$^1$, Nicholas~J.~Cooper$^4$, Anthony~J.~Milano$^1$, Michael~W.~Evans$^1$, Carolyn~C.~Porco$^5$, Joseph~N.~Spitale$^5$ \& John~W.~Weiss$^{5,6}$}

\affil{$^1$Department of Astronomy, Cornell University, Ithaca, NY 14853, USA.\\$^2$College of Engineering, Cornell University, Ithaca, NY 14853, USA.\\$^3$Laboratory for Atmospheric and Space Physics, University of Colorado at Boulder, 392 UCB, Boulder, CO 80309, USA.\\$^4$Astronomy Unit, Queen Mary University of London, Mile End Road, London E1 4NS, UK.\\$^5$CICLOPS, Space Science Institute, 4750 Walnut Street, Boulder, CO 80301, USA.\\$^6$Physics and Astronomy Department, Carleton College, 1 North College Street, Northfield, MN 55057, USA.}

\begin{abstract}
We report the discovery of several large ``propeller'' moons in the outer part of Saturn's A~ring, objects large enough to be followed over the 5-year duration of the \textit{Cassini} mission.  These are the first objects ever discovered that can be tracked as individual moons, but do not orbit in empty space.  We infer sizes up to 1--2~km for the unseen moonlets at the center of the propeller-shaped structures, though many structural and photometric properties of propeller structures remain unclear.  Finally, we demonstrate that some propellers undergo sustained non-keplerian orbit motion. 

\end{abstract}

\section{Introduction}

``Propeller'' structures in a planetary ring, named for their characteristic two-armed shape, occur as the disturbance caused by a disk-embedded moon is carried downstream, which is forward (backward) on the side facing toward (away from) the planet per Kepler's Third Law. Although no central moonlet has yet been directly resolved within a propeller, the observed structure allows us to infer both the locations and sizes of such moonlets.  Predictions of such structures in Saturn's rings \citep{SS00,SSD02,Seiss05} led to their discovery in \textit{Cassini} images \citep{Propellers06}, followed by the realization that they reside primarily in a relatively narrow band in the mid-A~ring \citep{Sremcevic07}.  Further observations identified three ``Propeller Belts'' between 127,000 and 132,000~km from Saturn's center, together containing 7000--8000 propeller moonlets with radius $R \gtrsim 0.15$~km, and exhibiting a steep power-law size distribution\fn{We refer to a cumulative (or integral) size distribution, of the form $N(R) \propto R^{-Q}$, where $N$ is the number of particles per unit area with radius greater than $R$.} of $Q \sim 5$ \citep{Propellers08}.  

\begin{figure*}[!t]
\begin{center}
\includegraphics[width=13.5cm,keepaspectratio=true]{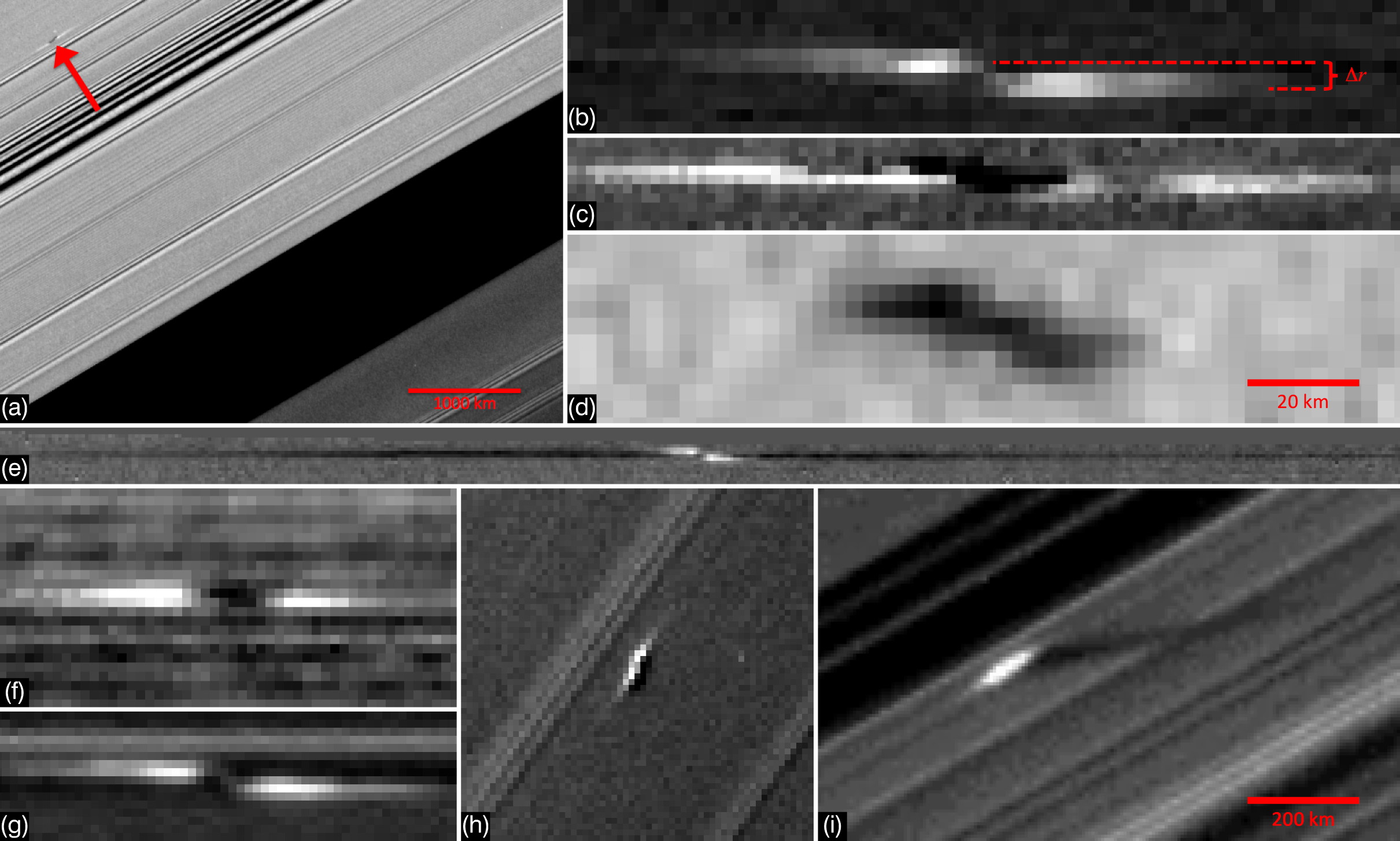}
\caption{Propellers as seen in selected \textit{Cassini} images.  Panel (a) shows a propeller in context of the Encke Gap and several density waves, while panels (h) and (i) show propellers casting shadows near the saturnian equinox (Section~\ref{shadows}).  The unseen moonlet is inferred to be at the center of the structure; its size is proportional to $\Delta r$ (Section~\ref{DeltaR_section}), the radial offset between the two azimuthally-aligned lobes, as illustrated in panel (b).  Panels (b), (c), and (d) show three views of the same propeller at the same scale, demonstrating how its appearance changes with viewing geometry.  Non-equinox views are on the lit (b,e,g) or unlit (a,c,d,f) face of the rings.  The scale bar in panel (d) also applies to panels (b), (c), (f), and (g).  The scale bar in panel (i) also applies to panels (e) and (h).  Image numbers, propeller nicknames, and apparition numbers are as follows:  (a) N1597800527, ``Santos-Dumont'', 081-081-A; (b) N1586641255, ``Bl\'eriot'', 064-255-A; (c) N1597791119, ``Bl\'eriot'', 081-054-A; (d) N1590907054, ``Bl\'eriot'', 070-015-A; (e) N1544842586, ``Bl\'eriot'', 035-299-A; (f) N1503243458, ``Wright'', 013-020-A; (g) N1586628622, ``Earhart'', 064-121-A; (h) N1620657077, ``Bl\'eriot'', 110-088-A; and (i) N1628846480, ``Earhart'', 116-008-A. 
\label{propeller_pics}}
\end{center}
\end{figure*}

We report here\fn{\citet{Sremcevic07} previously reported a single sighting of a propeller beyond the Encke Gap.} that a population of propellers also exists in the outer part of the A~ring, between the Encke Gap and the ring's outer edge (i.e., between 133,700 and 136,700~km from Saturn's center).  These ``\textit{trans}-Encke'' propellers are much less abundant but also much larger than those seen in the Propeller Belts (\Fig{}~\ref{propeller_pics}).  Because of this, several \textit{trans}-Encke propellers have been observed on multiple occasions with confidence that the same object is being viewed in each case, thus allowing conclusions to be drawn as to their persistence and orbital stability.  No propellers are clearly apparent between the Propeller Belts and the Encke Gap.

Each propeller observation\fn{Tables~\ref{prop_imtable} through~\ref{prop_table_deltar} contain details of the propeller observations that could not be included in the \textit{Astrophysical Journal Letters} version of this paper, due to lack of space.} is given an alphanumeric identifier, following the convention of \citet{Propellers08}.  However, propellers that are seen in multiple widely-separated apparitions are given nicknames that can serve to tie together the various observations.  To confirm the identity of objects in widely-separated apparitions, we check not only that the radial locations are the same but also that the longitudes are consistent (with residuals as specified in Section~3) with keplerian motion at that orbital radius. 

\section{Size and photometry \label{sizephot}}

\subsection{Size from radial offset \label{DeltaR_section}}
The size of the central moonlet, which is not directly seen, can be inferred from the radial offset $\Delta r$ between the leading and trailing azimuthally-aligned lobes of the propeller (\Fig{}~\ref{propeller_pics}b).  Initial simulations suggested that the radial separation between the density-depleted regions on either side of a moonlet is $\Delta r \sim 4 r_H$ \citep{Seiss05}, where $r_H$ is the Hill radius; interpretation of recent $N$-body simulations including mutual self-gravity and particle-size distributions is more complex due to moonlet accretion during the simulation, but the relationship $\Delta r \sim 4 r_H$ appears to hold \citep{LS09}.  Assuming that the density of the central moonlet is such that it fills its Roche lobe, and thus the moonlet's mean radius $\bar{R}$ is approximately 0.72 times its Hill radius \citep{PorcoSci07}, we then expect $\bar{R} \sim 0.18 \Delta r$.  

In the Propeller Belts, what appear morphologically to be the density-depleted regions within individual propellers unexpectedly appear bright against the background ring.  \citet{Sremcevic07} suggested brightening due to temporary liberation of regolith, while \citet{Anparsgw10} investigated the role that disruption of self-gravity wakes might play in brightening propeller structures.  Whatever the cause, brightened regions may blend, or even be primarily associated, with density-enhanced regions at larger values of $\Delta r$ (e.g.,~\citet{Sremcevic07} suggested $\Delta r \sim (9 \pm 1)r_H$ for density enhancements), so that $\bar{R} \lesssim 0.18 \Delta r$ 
should be regarded as an upper limit.  

Some of the largest propellers seen in the \textit{trans}-Encke region show, for the first time, both bright and dark components (\Fig{s}~\ref{propeller_pics}b through~\ref{propeller_pics}d).  For these, the density-enhanced and density-depleted regions can be clearly separated, under the assumption that the density-enhanced regions are brighter (darker) on the lit (unlit) face of the rings\fn{Brightness contrast for images of the rings' unlit face can be (and, for many viewing geometries of the A~ring, is) reversed, as regions with higher surface density are more opaque and thus appear darker.  And indeed, the general morphology of propellers seen on the unlit face appears to correspond in most cases to a contrast-reversed version of the morphology of propellers seen on the lit face.}.  Indeed, in some images of giant propellers the morphology is clearly that of density-enhanced regions, including the bright ribbon connecting ``Earhart's'' two bright lobes (\Fig{}~\ref{propeller_pics}g) and ``Bl\'eriot's'' bright lobes that are sometimes seen wrapped around dark wings (density-depleted regions) with a great deal more azimuthal extent (\Fig{}~\ref{propeller_pics}e).

\begin{figure*}[!t]
\begin{center}
\includegraphics[width=14cm,keepaspectratio=true]{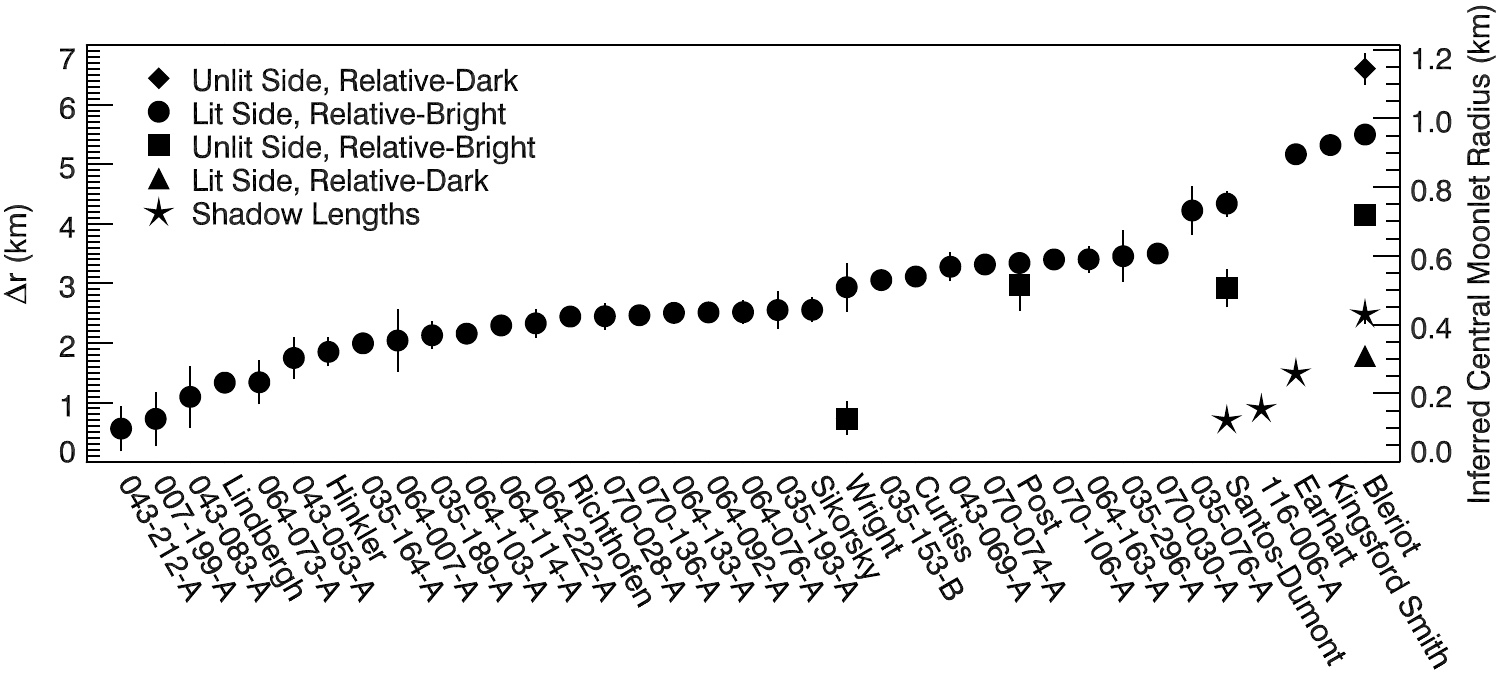}
\caption{Observed radial offsets $\Delta r$ for all propellers with measurement errors $< 0.7$~km.  Error bars (1-sigma) are given, but in many cases are smaller than the plotted symbol.  The right-hand axis gives the inferred radius of the central moonlet assuming internal density equal to the Roche critical density and $\Delta r \sim 4 r_H$; we emphasize that the latter assumption cannot be valid for all cases listed here (see Section~\ref{DeltaR_section}).  From shadow lengths we directly measure the vertical height of the propeller, which should be comparable to the moonlet radius (Section~\ref{shadows}).  For ``Bl\'eriot'', the only propeller with observed relative-dark components on the lit face of the rings or with observed relative-bright components on the unlit face of the rings, the radial offsets for lit-dark and unlit-bright do not agree, nor do unlit-dark and lit-bright, as might have been expected.  Furthermore, the four propellers observed on both the lit and unlit faces of the rings do not have a consistent ratio of lit-bright to unlit-bright.  
\label{prop_mass_plot}}
\end{center}
\end{figure*}

However, our measurements of $\Delta r$ in these cases (\Fig{}~\ref{prop_mass_plot}) show that further photometric and theoretical understanding is needed before the size of the central moonlet can be confidently obtained from $\Delta r$ to accuracy better than a factor of several (making the mass uncertain by about an order of magnitude).  We expected to find that relative-bright features on the lit face of the rings and relative-dark features on the unlit face would be the same (density-enhanced) components of the propeller, and similarly that relative-dark features on the lit face and relative-bright features on the unlit face would be the same (density-depleted) components, but in measured $\Delta r$ values for propeller ``Bl\'eriot'' (the only one for which all four aspects are seen) the radial separations all differ from each other.  Similarly, while relative-bright features on the lit face do have consistently larger values of $\Delta r$ than relative-bright features on the unlit face, as expected for (respectively) density-enhanced and density-depleted components, the ratios between the two are not consistent among the four propellers (``Bl\'eriot'', ``Post'', ``Santos-Dumont'', and ``Wright'') that have been observed on both the lit and unlit faces of the rings. 

The unexpected complexity of these observations indicates that the photometry of propellers is highly dependent on viewing geometry, perhaps including solar incidence angle and viewer's emission angle in addition to the more commonly-considered phase angle, and may also simply vary with time.  Such photometric variations may be due to the poorly-understood details of propeller structure and/or to differing particle-size properties between propeller and background. 

\subsection{Size from shadows \label{shadows}}

A saturnian equinox, which occurs every 14.7~yr, took place in 2009~August.  During this event, the Sun shone nearly edge-on to the ring plane, causing any vertical structure to cast long shadows.  Several propellers were observed during this period, along with their shadows (\Fig{s}~\ref{propeller_pics}h and~\ref{propeller_pics}i).  The azimuthally-distributed morphology of the shadows indicates that they were cast by the propeller structures as a whole, and not directly by the central moonlets.  However, the vertical extent of a propeller structure should be comparable to the size of its moonlet, as it is the moonlet's gravity that pulls material out of the ring plane (preliminary results from simulations corroborate this notion; M.~C.~Lewis, pers. comm.).  Shadow lengths indicate that ``Bl\'eriot's'' propeller has a vertical height of $0.43 \pm 0.03$~km above the ring, while other propellers have shadow-inferred vertical heights of 0.12~km (``Santos-Dumont''), 0.16~km (116-006-A) and 0.26~km (``Earhart'').  The main body of the A~ring has a vertical thickness on the order of a $\mathrm{few} \times 0.01$~km \citep{soirings,Hedman07,ColwellChapter09}. 

\subsection{Size distribution \label{sizedist}}

\begin{figure}[!t]
\begin{center}
\includegraphics[width=8.5cm,keepaspectratio=true]{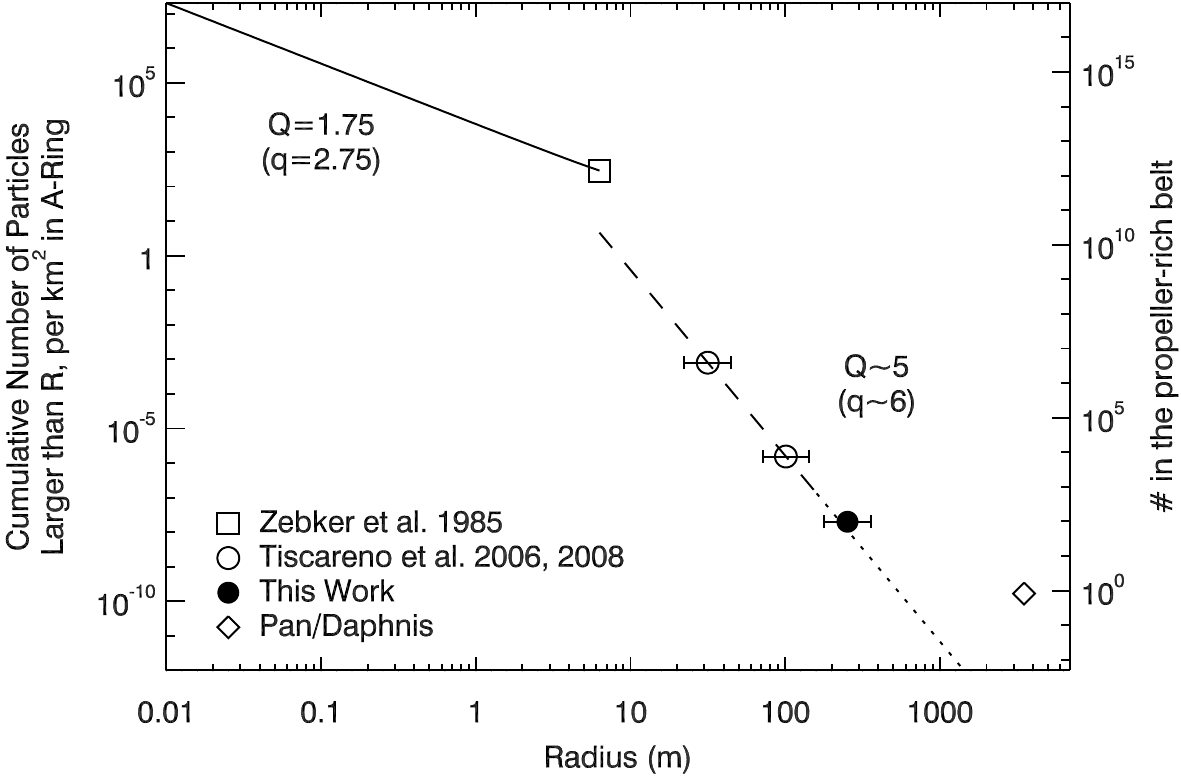}
\caption{Particle size distribution for Saturn's A Ring.  The solid line and open square denote the \textit{Voyager}~RSS size distribution for the ring continuum, converted to integral format from the solution for 2.14~$R_\mathrm{S} = 1$29,100~km given in Table~III of \citet{Zebker85}.  The open diamond is obtained from two known moons (Pan and Daphnis) of radius $\gtrsim 4$~km in the entire A~Ring.  The open circles denote values for the Propeller Belts, from \citet{Propellers06,Propellers08}.  The filled circle shows a result from this work for the \textit{trans}-Encke region.  The error bars reflect the systematic error due to model-dependency in the conversion from $\Delta r$ to moonlet size (the left-hand side of the error bar corresponds to $\Delta r \sim 4 r_H$, while the right-hand side of the error bar corresponds to $\Delta r \sim 8 r_H$, see Section~\ref{DeltaR_section}); it is important to note that data points will slide along the error bars in concert, preserving their relative positions and thus the inferred power law.  The change in linestyle from dashed to dotted is to indicate that the open circles and the filled circles represent data from different regions within the A~Ring, so that there is no reason to assume that they must fall on the same size-distribution curve, and furthermore that large propellers appear to be absent in the Propeller Belts while small propellers are absent in the \textit{trans}-Encke region (Section~\ref{sizedist}).  \label{prop_powerlaw}}
\end{center}
\end{figure}

The propeller-belt and the giant-propeller populations may in fact have very similar size-distribution curves.  The complete azimuthal scan from Orbit~35 found 19 propellers between 133,700~and 135,000~km with $\Delta r \gtrsim 2$~km, and the partial azimuthal scan (65\%-complete) from Orbit~64 found 16, both of which give a surface density\fn{We note that the surface density for \textit{trans}-Encke propellers reported in \Fig{}~8 of \citet{Propellers08} was too low due to an erroneous assumption that the azimuthal scan from Orbit~13 was complete.} of $2 \times 10^{-8}$~km$^{-2}$.  
As shown in \Fig{}~\ref{prop_powerlaw}, this is the same number that one gets by extrapolating from the Propeller Belts' surface density and power law $Q \sim 5$ \citep{Propellers08}. 

\begin{figure*}[!t]
\begin{center}
\includegraphics[width=15cm,keepaspectratio=true]{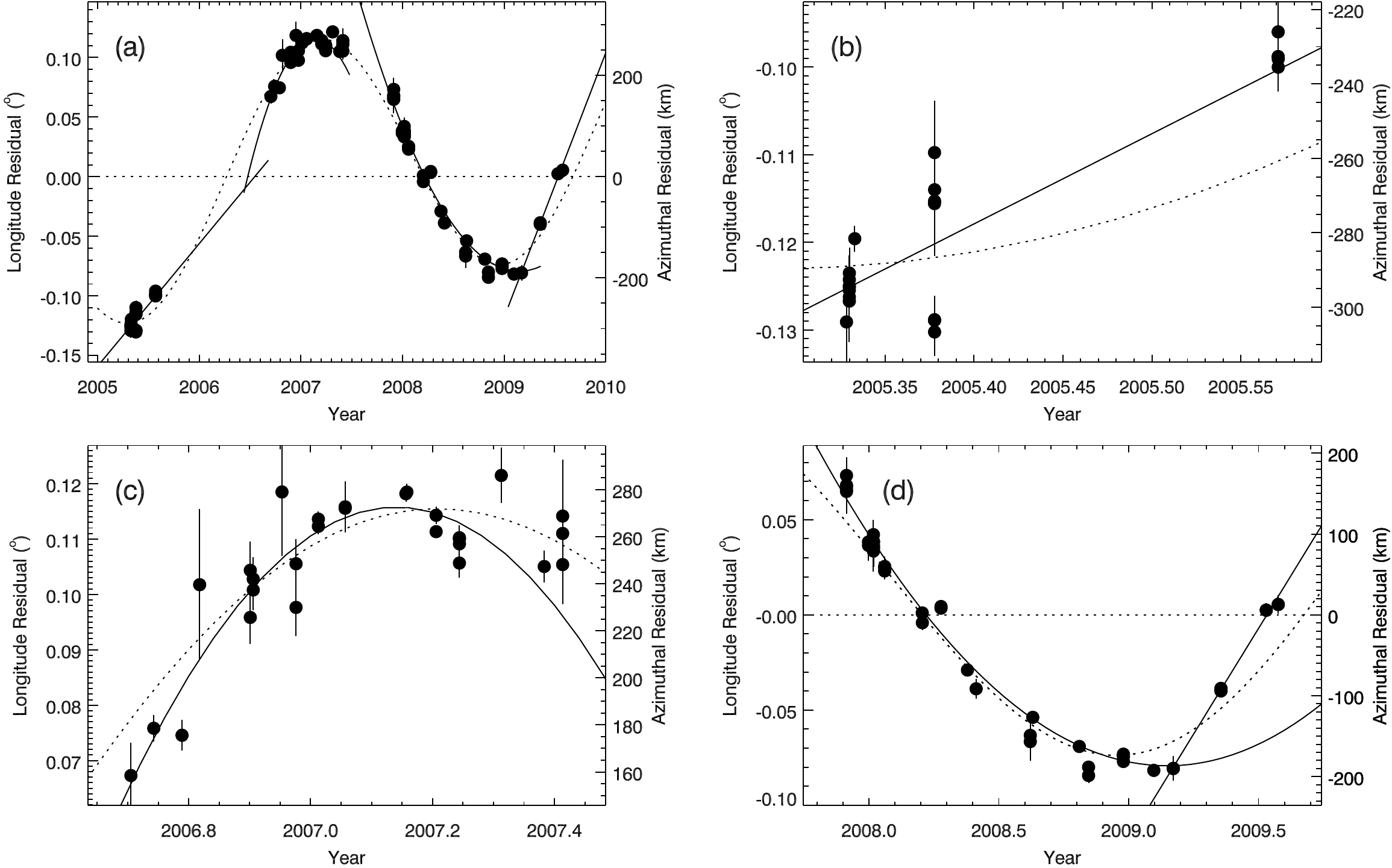}
\caption{Observed longitude of the propeller ``Bl\'eriot'' over 4~years, with a linear trend (616.7819329$^\circ$/day) subtracted off.  Only data points with measurement errors $\sigma < 0.01^\circ$ are shown.  Error bars (1-sigma) are given, but in many cases are smaller than the plotting symbol.  Panel (a) shows all the data, while panels (b), (c), and (d) contain subsets of the data shown in greater detail.  The residuals to the linear trend (horizontal dotted line) are less than $\pm 300$~km, but are clearly not randomly distributed.  The dotted line indicates a linear-plus-sinusoidal fit to all the data, with an amplitude of $0.11^\circ$ and a period of 3.68~yr.  The solid lines indicate piecewise quadratic fits, corresponding to a constant drift in semimajor axis; in particular, the data from mid-2006 to early-2007 (panel~c) are fit by a linear trend with a constant acceleration of -0.0096$''$/day$^2$ ($\dot{a} = +0.11$~km/yr), while the data from late-2007 to early-2009 (panel~d) are fit by a linear trend with a constant acceleration of +0.0023$''$/day$^2$ ($\dot{a} = -0.04$~km/yr). 
\label{bleriot_orbit}}
\end{center}
\end{figure*}

\begin{table*}[!t]
\begin{center}
\caption{Orbit fits for \textit{trans}-Encke propellers
\label{propeller_fits}}
\begin{scriptsize}
\begin{tabular}{l r @{.} l r @{.} l c c c c c }
\hline
\hline
\vspace{-0.1cm}
& \multicolumn{2}{c}{} & \multicolumn{2}{c}{} & Longitude & & & \multicolumn{2}{c}{Rms deviation} \\
Nickname & \multicolumn{2}{c}{$n$, $^\circ$/day$^a$} & \multicolumn{2}{c}{$a$, km$^a$} & at epoch$^b$ & \# images$^c$ & Time interval & in km & in longitude \\
\hline
Earhart & 624&529897(2) & 133797&8401(3) & 57.85$^\circ$ &  3 & 2006--2009 (2.7 yr) & 730 & 0.31$^\circ$ \\
Post & 624&4867(3) & 133803&99(4) & 58.09$^\circ$ &  3 & 2006--2008 (1.7 yr) & 12 & 0.01$^\circ$ \\
Sikorsky & 623&917736(1) & 133885&0475(2) & 70.37$^\circ$ &  3 & 2005--2008 (3.1 yr) & 230 & 0.10$^\circ$ \\
Curtiss & 623&7473 & 133909&36 & 210.04$^\circ$ &  2 & 2006--2008 (1.7 yr) & & \\
Lindbergh & 623&3176(2) & 133970&69(2) & 112.08$^\circ$ &  3 & 2005--2008 (3.0 yr) & 71 & 0.03$^\circ$ \\
Wright & 622&5527 & 134080&03 & 251.85$^\circ$ &  2 & 2005--2006 (1.3 yr) & & \\
Kingsford Smith & 620&761649(2) & 134336&9350(3) & 202.44$^\circ$ &  4 & 2005--2008 (2.9 yr) & 670 & 0.28$^\circ$ \\
Hinkler & 619&80519(1) & 134474&639(2) & 58.85$^\circ$ &  3 & 2006--2008 (1.3 yr) & 360 & 0.15$^\circ$ \\
Santos-Dumont & 619&458729(1) & 134524&6067(2) & 324.11$^\circ$ &  9 & 2005--2009 (4.3 yr) & 670 & 0.28$^\circ$ \\
Richthofen & 617&7011 & 134778&83 & 122.90$^\circ$ & 2 & 2006--2007 (0.3 yr) & & \\
Bl\'eriot & 616&7819329(6) & 134912&24521(8) & 193.65$^\circ$ & 89 & 2005--2009 (4.2 yr) & 210 & 0.09$^\circ$ \\
\hline
\multicolumn{10}{l}{$^a$ Formal error estimates, shown in parentheses for the last digit, are for the best-fit linear trend in longitude. They are}\\
\multicolumn{10}{l}{much smaller than the rms deviations in longitude, given in the right-hand column.} \\
\multicolumn{10}{l}{$^b$ Epoch is 2007~January~1 at 12:00:00~UTC (JD~1782806.0).  All orbit fits assume $e=0$ and $i=0$. }\\
\multicolumn{10}{l}{$^c$ Not including images of insufficient quality to include in the orbit fit.} \\
\end{tabular}
\end{scriptsize}
\end{center}
\end{table*}

However, if the two populations do have the same size-distribution slope and intercept, they are likely truncated so that they do not actually overlap.  The same radial scans that discovered the Propeller Belts \citep{Propellers08} clearly showed no small propellers elsewhere in the A~ring, including the \textit{trans}-Encke region, so it would appear that only giant propellers occur in the latter.  On the other hand, while fewer high-resolution (better than 3~km/px) images have been taken of the Propeller Belts, compared with the \textit{trans}-Encke region, the ratio is only a factor of several, which makes it likely (but not conclusively proven) that giant propellers are missing in the Propeller Belts.  Even the largest propellers observed in the Propeller Belts have $\Delta r < 1.3$~km \citep{Propellers08}, while nearly all observed \textit{trans}-Encke propellers have $\Delta r$ larger than this value (\Fig{}~\ref{prop_mass_plot}). 

\section{The orbital evolution of ``Bl\'eriot'' \label{OrbitalEvolution}}

At least 11~propellers have been seen at multiple widely-separated instances, but ``Bl\'eriot'' is of particular interest as the largest and most frequently detected (\Fig{s}~\ref{propeller_pics}b, \ref{propeller_pics}c, \ref{propeller_pics}d, \ref{propeller_pics}e, and~\ref{propeller_pics}h).  It has appeared in more than one hundred separate \textit{Cassini}~ISS images spanning a period of four years, and was serendipitously detected once in a stellar occultation observed by the \textit{Cassini}~UVIS instrument \citep{ColwellEPSC08}. 

Analysis of the orbit of ``Bl\'eriot'' confirms that it is both long-lived and reasonably well-characterized by a keplerian path.  As \Fig{}~\ref{bleriot_orbit} shows, a linear fit to the longitude with time (corresponding to a circular orbit) results in residuals of $\pm 300$~km (0.13$^\circ$ longitude).  However, those residuals are some 50~times greater than the mean measurement errors; furthermore, the residuals are far from randomly distributed, with a strong indication of coherent motions superposed atop the linear trend.  Analysis of the orbits of other propellers, though less well-sampled in time, similarly shows variations in longitude of up to 1000~km (Table~\ref{propeller_fits}), easily small enough to confirm that a single object is being observed but much larger than the measurement errors. 

After a linear trend, the next-simplest functional form available is a quadratic, which corresponds to a constant angular acceleration, or (equivalently) a linear drift of the semimajor axis with time.  A global quadratic fit to all data points does hardly better than the global linear fit and is not shown in \Fig{}~\ref{bleriot_orbit}.  

Two possible functional forms for the non-keplerian motion are 1) a linear trend plus a sinusoidal variation of the longitude with time, which is shown by the dotted line in \Fig{}~\ref{bleriot_orbit}, and 2) a piecewise fit carried out by grouping the data into four segments and fitting each segment to a linear or quadratic trend, shown by the solid lines in \Fig{}~\ref{bleriot_orbit}.  The most likely physical mechanism for the former is some kind of resonant interaction, perhaps with one of the many larger moons in or near the rings, although no resonance has yet been identified that might plausibly explain ``Bl\'eriot's'' motions.  The most likely physical mechanism for the latter is that ``Bl\'eriot'' periodically suffers collisions \citep{LS09} that jostle its orbit onto a new one, but that in between such kicks its semimajor axis drifts linearly due to gravitational and/or collisional interactions with the disk.  

Continued investigation of the non-keplerian motion of propeller moons is ongoing. 

\acknowledgements This paper is dedicated to the memory of Kevin~Beurle.  We thank B.~M.~Byington for help with image processing.  We thank the Cassini project and the Cassini Imaging Team.  M.S.T. acknowledges funding from the NASA Cassini Data Analysis program (NNX08AQ72G and NNX10AG67G).

\renewcommand{\thetable}{S\arabic{table}}
\setcounter{table}{0}

\begin{table*}
\begin{center}
\caption{Observing information for images used in this paper. \label{prop_imtable}}
\begin{scriptsize}
\hspace{-0.5cm}
\begin{tabular} { c c c c c c c c c }
\hline
\hline
 & & \# of & & Incidence & Emission & Phase & Radial & Azimuthal \\
Orbit & Image Identifier & Images & Date & Angle$^b$ & Angle$^b$ & Angle & Resolution$^c$ & Resolution$^c$ \\
\hline
007 & N1493569190 & 1 & 2005-120 & 111.9$^\circ$ & 109.6$^\circ$  &  40.2$^\circ$  &   8.6 &  25.1 \\
007 & N1493619126 -- 42136 & 20 & 2005-121 & 111.8$^\circ$ & 111.0$^\circ$  &  33.6$^\circ$  &   6.9 &  18 --  20 \\
007 & N1493715564 -- 24205 & 3 & 2005-122 & 111.8$^\circ$ & 115.2$^\circ$  &  12.2$^\circ$  &   3 --   5 &   3 --   6 \\
008 & N1495108987 -- 35027 & 17 & 2005-138 & 111.7$^\circ$ & 109.6$^\circ$  &  41.7$^\circ$  &   8.9 &  25 --  28 \\
009 & N1496871927 -- 2034 & 3 & 2005-158 & 111.5$^\circ$ & 109.4$^\circ$  &  18.5$^\circ$  &   3.8 &   6.6 \\
012 & N1501228513 -- 9785 & 4 & 2005-209 & 110.9$^\circ$ & 107.4$^\circ$  &  55.7$^\circ$  &  13.8 &  19 --  24 \\
013 & N1503243458 & 1 & 2005-232 & 110.7$^\circ$ &  52.1$^\circ$  & 162.3$^\circ$  &   1.3 &   1.1 \\
028 & N1537022412 & 1 & 2006-258 & 105.9$^\circ$ &  75.0$^\circ$  & 175.8$^\circ$  &  13.0 &  49.6 \\
029 & N1538205758 & 1 & 2006-271 & 105.7$^\circ$ &  58.7$^\circ$  & 161.0$^\circ$  &  10.1 &  19.5 \\
030 & N1539660972 -- 8180 & 2 & 2006-289 & 105.4$^\circ$ &  49.0$^\circ$  & 151.2$^\circ$  &  10.6 &  16.1 \\
031 & N1540581552 & 1 & 2006-299 & 105.3$^\circ$ & 106.6$^\circ$  & 117.8$^\circ$  &   5.9 &  10.7 \\
033 & N1542569654 -- 70226 & 2 & 2006-322 & 105.0$^\circ$ &  91.9$^\circ$  & 147.9$^\circ$  &   7.2 & 150 -- 174 \\
033 & N1543202188 -- 695 & 2 & 2006-329 & 104.9$^\circ$ &  59.8$^\circ$  & 160.5$^\circ$  &  10.2 &  20.4 \\
034 & N1543354660 -- 5569 & 2 & 2006-331 & 104.9$^\circ$ &  69.6$^\circ$  & 159.0$^\circ$  &  10.2 &  29.2 \\
035 & N1544813989 -- 43793 & 20 & 2006-348/349 & 104.6$^\circ$ & 117$^\circ$ -- 150$^\circ$ &  27$^\circ$ --  64$^\circ$ &   2.7 &   2 --   5 \\
036 & N1545570886 -- 1332 & 2 & 2006-357 & 104.5$^\circ$ &  54.9$^\circ$  & 159.2$^\circ$  &  11.8 &  20.6 \\
036 & N1546726425 -- 798 & 2 & 2007-005 & 104.3$^\circ$ &  35.3$^\circ$  & 133.6$^\circ$  &  10.4 &  12.8 \\
037 & N1548113882 -- 946 & 2 & 2007-021 & 104.1$^\circ$ &  34.0$^\circ$  & 110.2$^\circ$  &  10.7 &  10.4 \\
039 & N1551257093 -- 309107 & 2 & 2007-058 & 103.6$^\circ$ &  34.8$^\circ$  & 104.9$^\circ$  &  10.2 &  12.4 \\
041 & N1552822117 -- 477 & 2 & 2007-076 & 103.3$^\circ$ &  35.5$^\circ$  & 108.8$^\circ$  &  10.6 &  13.1 \\
041 & N1554027273 -- 8323 & 4 & 2007-090 & 103.1$^\circ$ &  51.8$^\circ$  &  82.2$^\circ$  &  11.9 &  19.2 \\
042 & N1554731052 & 1 & 2007-098 & 103.0$^\circ$ & 118.9$^\circ$  & 128.2$^\circ$  &   2.7 &   5.6 \\
043 & N1556183373 -- 202155 & 6 & 2007-115 & 102.8$^\circ$ & 123$^\circ$ -- 138$^\circ$ &  21$^\circ$ --  43$^\circ$ &   3.1 --   4.2 &   2.6 --   4.5 \\
045 & N1558418169 & 1 & 2007-141 & 102.4$^\circ$ &  71.0$^\circ$  &  72.4$^\circ$  &  13.7 &  23.7 \\
045 & N1559371611 -- 77 & 3 & 2007-152 & 102.2$^\circ$ &  83.4$^\circ$  &  51.5$^\circ$  &  12.5 & 105.4 \\
053 & N1575154715 -- 65751 & 9 & 2007-334/335 &  99.5$^\circ$ &  80.4$^\circ$  &  52.2$^\circ$  &   8.7 &  49 --  52 \\
054 & N1577141652 -- 95 & 2 & 2007-356 &  99.2$^\circ$ &  75.9$^\circ$  &  23.9$^\circ$  &  11.9 &  33.7 \\
055 & N1577829421 -- 30077 & 3 & 2007-365 &  99.0$^\circ$ &  57.3$^\circ$  &  64.8$^\circ$  &   9.6 &  17.7 \\
055 & N1578422585 -- 9943 & 12 & 2008-007 &  98.9$^\circ$ &  78.1$^\circ$  &  21.8$^\circ$  &   9.9 &  47.6 --  49.0 \\
056 & N1579166576 & 1 & 2008-016 &  98.8$^\circ$ & 123.8$^\circ$  &  66.2$^\circ$  &   2.5 &   4.5 \\
057 & N1579792597 -- 3315 & 3 & 2008-023 &  98.7$^\circ$ &  62.7$^\circ$  &  42.7$^\circ$  &  10.2 &  22.2 \\
061 & N1584357723 -- 73623 & 2 & 2008-067 &  97.9$^\circ$ &  77.8$^\circ$  &  20.9$^\circ$  &  12.4 &  14.7 \\
064 & N1586618575 -- 41255 & 17 & 2008-102 &  97.5$^\circ$ & 141$^\circ$ -- 175$^\circ$ &  49$^\circ$ -- 103$^\circ$ &   1.7 --   2.8 &   1.8 --   3.4 \\
065 & N1587553446 -- 4566 & 5 & 2008-113 &  97.3$^\circ$ & 104.5$^\circ$  &  14.5$^\circ$  &   5.5 &  21.8 \\
068 & N1589849522 & 1 & 2008-139 &  96.9$^\circ$ & 112.5$^\circ$  &  37.3$^\circ$  &   5.3 &   8.4 \\
070 & N1590907054 -- 8480 & 2 & 2008-152 &  96.7$^\circ$ &  56.6$^\circ$  &  42.5$^\circ$  &   6.3 &   7 --  11 \\
070 & N1591067764 -- 85510 & 5 & 2008-154 &  96.7$^\circ$ & 140$^\circ$ -- 154$^\circ$ &  71$^\circ$ --  82$^\circ$ &   2.3 &   2.1 \\
071 & N1591525824 & 1 & 2008-159 &  96.6$^\circ$ &  58.8$^\circ$  &  37.9$^\circ$  &   6.3 &  11.1 \\
080 & N1597462656 & 1 & 2008-228 &  95.6$^\circ$ &  87.1$^\circ$  &  26.5$^\circ$ &   7.4 &  58.4 \\
080 & N1597487541 -- 8439 & 4 & 2008-228 &  95.6$^\circ$ &  83.4$^\circ$  &  16.7$^\circ$ &   7.7 &  20 --  60 \\
081 & N1597775567 -- 800527 & 4 & 2008-231/232 &  95.5$^\circ$ &  25$^\circ$ --  45$^\circ$ &  94$^\circ$ -- 105$^\circ$ &   3.0 &   2.7 \\
090 & N1603444112 & 1 & 2008-297 &  94.5$^\circ$ &  37.0$^\circ$  &  58.9$^\circ$  &   4.4 &   5.5 \\
092 & N1604569991 -- 70034 & 2 & 2008-310 &  94.3$^\circ$ &  69.3$^\circ$  &  32.7$^\circ$  &   8.0 &  12.7 \\
098 & N1608800086 -- 341 & 3 & 2008-359 &  93.5$^\circ$ &  24.8$^\circ$  &  69.9$^\circ$  &   5.3 &   4.9 \\
102 & N1612496648 & 1 & 2009-036 &  92.9$^\circ$ & 128.6$^\circ$  &  59.0$^\circ$  &   9.5 &   7.0 \\
105 & N1614861737 & 1 & 2009-063 &  92.5$^\circ$ &  66.5$^\circ$  &  34.5$^\circ$  &   7.1 &  17.8 \\
110 & N1620656882 -- 7077 & 2 & 2009-130 &  91.4$^\circ$ & 117.5$^\circ$  &  59.9$^\circ$  &   5.6 &  12.2 \\
114 & N1626159520 & 1 & 2009-194 &  90.4$^\circ$ &  41.8$^\circ$  &  77.3$^\circ$ &   9.7 &   7.2 \\
114 & N1626320702 & 1 & 2009-196 &  90.4$^\circ$ &  49.7$^\circ$  &  89.4$^\circ$ &   11.5 &   13.1 \\
115 & N1627613635 & 1 & 2009-211 &  90.2$^\circ$ &  61.2$^\circ$  &  99.1$^\circ$  &  10.1 &  20.8 \\
116 & N1628845563 -- 6513 & 7 & 2009-225 &  90.0$^\circ$ &  72.0$^\circ$  &  82$^\circ$ --  92$^\circ$ &   8 --  11 &   9 --  13 \\
\hline
\multicolumn{9}{l}{$^a$ For lines referring to multiple images, variation in each parameter is in the last significant figure. }\\
\multicolumn{9}{l}{$^b$ Measured from the direction of Saturn's north pole (ring-plane normal), so that angles $>90^\circ$ denote the southern} \\
\multicolumn{9}{l}{hemisphere. }\\
\multicolumn{9}{l}{$^c$ In km/pixel.}\\
\end{tabular}
\end{scriptsize}
\end{center}
\end{table*}

\begin{table*}[!t]
\begin{center}
\caption{Fitted sizes for multiply-observed \textit{trans}-Encke propellers.
\label{prop_mass_table}}
\begin{scriptsize}
\begin{tabular}{l c c r @{$\pm$} l c c c }
\hline
\hline
Name$^{a,b}$ & Lit/Unlit & B/D$^c$ & \multicolumn{2}{c}{$\Delta r$, km} & Inferred $\bar{R}$, km$^d$ & \# images & Reduced $\chi^2$ \\
\hline
064-103-A & Lit & B & 2.16 & 0.12 & $0.37 \pm 0.02$ &  2 & 13.73 \\
Bl\'eriot & Lit & B & 5.50 & 0.05 & $0.95 \pm 0.01$ &  6 &  9.99 \\
Bl\'eriot & Lit & D & 1.78 & 0.07 & $0.31 \pm 0.01$ &  7 &  3.75 \\
Bl\'eriot & Unlit & B & 4.15 & 0.18 & $0.72 \pm 0.03$ &  6 &  2.24 \\
Bl\'eriot & Unlit & D & 6.60 & 0.27 & $1.14 \pm 0.05$ &  3 &  0.58 \\
Curtiss & Lit & B & 3.11 & 0.10 & $0.54 \pm 0.02$ &  1 \\
Earhart & Lit & B & 5.17 & 0.11 & $0.90 \pm 0.02$ &  2 & 19.58 \\
Hinkler & Lit & B & 1.85 & 0.24 & $0.32 \pm 0.04$ &  2 &  0.00 \\
Kingsford Smith & Lit & B & 5.32 & 0.10 & $0.92 \pm 0.02$ &  2 &  0.04 \\
Lindbergh & Lit & B & 1.33 & 0.12 & $0.23 \pm 0.02$ &  2 &  0.15 \\
Post & Lit & B & 3.34 & 0.07 & $0.58 \pm 0.01$ &  3 & 24.63 \\
Post & Unlit & B & 2.97 & 0.43 & $0.51 \pm 0.07$ &  1 \\
Richthofen & Lit & B & 2.44 & 0.16 & $0.42 \pm 0.03$ &  2 &  0.78 \\
Santos-Dumont & Lit & B & 4.34 & 0.22 & $0.75 \pm 0.04$ &  3 & 22.27 \\
Santos-Dumont & Unlit & B & 2.92 & 0.32 & $0.51 \pm 0.06$ &  1 \\
Sikorsky & Lit & B & 2.56 & 0.21 & $0.44 \pm 0.04$ &  2 &  0.39 \\
Wright & Lit & B & 2.94 & 0.41 & $0.51 \pm 0.07$ &  1 \\
Wright & Unlit & B & 0.72 & 0.11 & $0.12 \pm 0.02$ &  1 \\
Wright & Unlit & D & 0.73 & 0.28 & $0.13 \pm 0.05$ &  1 \\
\hline
\multicolumn{8}{l}{$^a$ Alphanumeric identifier for a single apparition (see Table~\ref{prop_table_deltar}), or nickname used to tie} \\
\multicolumn{8}{l}{together multiple apparitions of the same object.} \\
\multicolumn{8}{l}{$^b$ Propellers seen only once are not listed here, as their measured $\Delta r$ values are already given } \\
\multicolumn{8}{l}{in Table~\ref{prop_table_deltar} and \Fig{}~\ref{prop_mass_plot}.} \\
\multicolumn{8}{l}{$^c$  Relative-bright (B) or relative-dark (D), with respect to image background.} \\
\multicolumn{8}{l}{$^d$ Mean radius of moonlet assuming internal density equal to the Roche critical density and } \\
\multicolumn{8}{l}{$\Delta r \sim 4 r_H$; we emphasize that the latter assumption cannot be valid for all cases listed here} \\
\multicolumn{8}{l}{(see text).} \\
\end{tabular}
\end{scriptsize}
\end{center}
\end{table*}

\begin{table*}[!t]
\begin{center}
\caption{Vertical heights of propellers, inferred from near-equinox shadows. 
\label{prop_shadow_table}}
\begin{scriptsize}
\begin{tabular}{l c c r @{$\pm$} l c c }
\hline
\hline
& & Solar & \multicolumn{2}{c}{Shadow} & Inferred & \\
& & Incidence & \multicolumn{2}{c}{Length,} & Obstacle & \\
Name$^a$ & Image & Angle$^b$ ($^\circ$) & \multicolumn{2}{c}{km} & Height, km & Nickname$^a$ \\
\hline
110-087-A & N1620656882 & 91.42 &  19 &  7 & $0.47 \pm 0.17$ & Bleriot\\
110-088-A & N1620657077 & 91.42 &  20 &  7 & $0.49 \pm 0.17$ & Bleriot\\
\hline
114-001-A & N1626159520 & 90.44 &  55 &  7 & $0.424\pm0.057$ & Bleriot\\
114-015-A & N1626320702 & 90.41 &  49 & 13 & $0.353 \pm 0.092$ & Bleriot\\
\hline
116-004-A & N1628845780 & 89.96 & 221 &  10 & $0.136 \pm 0.006$ & Santos-Dumont\\
116-005-A & N1628845813 & 89.96 & 184 &  10 & $0.113 \pm 0.006$ & Santos-Dumont\\
116-006-A & N1628846210 & 89.96 & 257 & 10 & $0.159 \pm 0.006$ & 116-006-A\\
116-007-A & N1628846243 & 89.96 & 252 & 10 & $0.156 \pm 0.006$ & 116-006-A\\
116-008-A & N1628846480 & 89.96 & 426 &  9 & $0.263 \pm 0.006$ & Earhart\\
116-009-A & N1628846513 & 89.96 & 419 &  9 & $0.259 \pm 0.006$ & Earhart\\
\hline
\multicolumn{7}{l}{$^a$ As in Table~\ref{prop_mass_table}.} \\
\multicolumn{7}{l}{$^b$ Measured from Saturn's north pole.} \\
\end{tabular}
\end{scriptsize}
\end{center}
\end{table*}

\clearpage
\begin{table*}[!h]
\begin{center}
\caption{Fitted sizes for \textit{trans}-Encke propellers.
\label{prop_table_deltar}}
\begin{scriptsize}
\begin{tabular}{ l c c c r @{$\pm$} l r @{$\pm$} l c }
\hline
\hline
Name$^a$ & Image & B/D$^b$ & [ line, sample ] & \multicolumn{2}{c}{$\Delta \ell$ (km)$^c$} & \multicolumn{2}{c}{$\Delta r$ (km)} & Match/Nickname$^d$\\
\hline
007-030-A & N1493619126 & B & [  217.2,1017.1 ] &  51.2 &   2.8 &   6.5 &   0.9 & Bl\'eriot\\
007-031-A & N1493619321 & B & [  247.4, 857.4 ] &  53.4 &   2.7 &   8.4 &   0.7 & Bl\'eriot\\
007-031-A &      "      & D &      "      & \multicolumn{2}{c}{} &   0.817 &  0.868 & Bl\'eriot\\
007-032-A & N1493619516 & B & [  269.3, 701.2 ] &  48.8 &   3.7 &   7.4 &   1.0 & Bl\'eriot\\
007-032-A &      "      & D &      "      & \multicolumn{2}{c}{} &   1.488 &  0.696 & Bl\'eriot\\
007-033-A & N1493619711 & B & [  278.2, 540.6 ] &  54.3 &   2.9 &   7.7 &   0.9 & Bl\'eriot\\
007-033-A &      "      & D &      "      & \multicolumn{2}{c}{} &   1.988 &  0.658 & Bl\'eriot\\
007-034-A & N1493619906 & B & [  277.9, 386.3 ] &  45.9 &   3.9 &   5.6 &   1.8 & Bl\'eriot\\
007-034-A &      "      & D &      "      & \multicolumn{2}{c}{} &   0.976 &  0.660 & Bl\'eriot\\
007-035-A & N1493620101 & B & [  267.1, 227.0 ] &  48.4 &   7.6 &   5.9 &   2.0 & Bl\'eriot\\
007-035-A &      "      & D &      "      & \multicolumn{2}{c}{} &   1.419 &  0.497 & Bl\'eriot\\
007-036-A & N1493620296 & B & [  244.9,  68.8 ] &  47.2 &   4.2 &   4.8 &   1.2 & Bl\'eriot\\
007-036-A &      "      & D &      "      & \multicolumn{2}{c}{} &   1.695 &  0.670 & Bl\'eriot\\
\hline
007-087-A & N1493715564 & D & [  318.8, 475.4 ] & \multicolumn{2}{c}{} &   2.253 &  0.125 & Bl\'eriot\\
007-173-A & N1493721598 & B & [  324.5, 924.0 ] &  33.1 &   1.5 &   1.9 &   0.8 & \\
007-199-A & N1493724205 & B & [  121.4, 670.5 ] &  31.0 &   0.8 &   0.7 &   0.5 & \\
\hline
008-159-A & N1495131307 & B & [  678.8, 814.2 ] &  40.9 &   8.5 &   6.1 &   3.4 & Bl\'eriot\\
008-160-A & N1495131555 & B & [  701.8, 659.8 ] &  46.8 &   5.9 &   5.4 &   1.9 & Bl\'eriot\\
008-161-A & N1495131803 & B & [  709.3, 502.0 ] &  53.8 &   5.0 &   3.9 &   1.5 & Bl\'eriot\\
008-162-A & N1495132051 & B & [  703.7, 350.0 ] &  50.4 &   5.9 &   5.8 &   2.0 & Bl\'eriot\\
008-163-A & N1495132299 & B & [  683.4, 193.8 ] &  50.1 &   6.5 &   6.3 &   2.5 & Bl\'eriot\\
008-164-A & N1495132547 & B & [  649.2,  33.3 ] &  67.7 &   5.1 &   1.7 &   1.7 & Bl\'eriot\\
\hline
009-023-A & N1496871927 & B & [  760.7, 590.4 ] &  31.0 &   1.9 &   7.7 &   0.7 & Santos-Dumont\\
009-025-A & N1496872000 & B & [  763.8, 585.6 ] &  28.3 &   2.2 &   6.2 &   0.6 & Santos-Dumont\\
009-026-A & N1496872034 & B & [  763.2, 584.6 ] &  20.4 &   2.7 &   8.1 &   0.9 & Santos-Dumont\\
\hline
012-049-A & N1501228513 & B & [  977.0, 122.7 ] &  68.9 &   5.1 &  15.7 &   1.9 & Bl\'eriot\\
012-050-A & N1501228937 & B & [  908.7,  90.2 ] &  48.1 &   9.4 &  14.0 &   2.8 & Bl\'eriot\\
012-051-A & N1501229361 & B & [  834.0,  61.3 ] &  63.7 &   5.4 &  14.0 &   2.1 & Bl\'eriot\\
012-052-A & N1501229785 & B & [  756.7,  28.8 ] &  61.6 &   6.7 &  14.6 &   2.0 & Bl\'eriot\\
\hline
013-020-A & N1503243458 & D & [  221.3, 425.3 ] &   3.4 &   0.5 &   0.7 &   0.3 & Wright\\
013-020-A &      "      & B & [  221.6, 426.0 ] &  22.1 &   0.4 &   0.7 &   0.1 & Wright\\
\hline
031-017-A & N1540581552 & B & [  417.0,  57.1 ] &  54.9 &   2.0 &   8.7 &   0.6 & Bl\'eriot\\
\hline
033-070-A & N1543202188 & B & [  617.2,  77.6 ] &  97.6 &   5.8 &   8.5 &   1.8 & Bl\'eriot\\
033-071-A & N1543202695 & B & [  199.5,  51.4 ] & 130.1 &   9.5 &   1.1 &   2.4 & Bl\'eriot\\
\hline
034-009-A & N1543354660 & B & [  566.8, 673.0 ] & 119.4 &   6.4 &   9.6 &   2.8 & Bl\'eriot\\
034-010-A & N1543355569 & B & [   44.0, 594.4 ] & 126.6 &   9.7 &   2.3 &   2.0 & Bl\'eriot\\
\hline
035-028-A & N1544813989 & B & [  790.8, 322.6 ] &  33.6 &   0.4 &   5.3 &   0.1 & Kingsford Smith\\
035-067-A & N1544818130 & B & [  966.8, 787.1 ] &  13.9 &   0.4 &   2.5 &   0.2 & Richthofen\\
035-076-A & N1544819124 & B & [  580.3,  27.9 ] &  14.7 &   1.4 &   4.2 &   0.4 & \\
035-094-A & N1544820869 & B & [  629.5, 842.8 ] &  17.7 &   0.2 &   3.1 &   0.1 & Curtiss\\
035-130-A & N1544824600 & B & [  919.6, 137.0 ] &  11.6 &   0.8 &   2.7 &   0.7 & \\
035-153-A & N1544826831 & B & [  651.7, 604.0 ] &  12.6 &   0.6 &   1.4 &   0.3 & Lindbergh\\
035-153-B &      "      & B & [  942.4, 687.6 ] &   9.0 &   0.3 &   3.1 &   0.2 & \\
035-164-A & N1544828015 & B & [  646.5, 487.7 ] &  13.3 &   0.3 &   2.0 &   0.2 & \\
035-189-A & N1544830440 & B & [  635.6, 298.9 ] &  20.6 &   0.4 &   2.1 &   0.2 & \\
035-193-A & N1544830828 & B & [  805.7, 282.4 ] &  14.0 &   0.4 &   2.6 &   0.3 & \\
035-206-A & N1544832207 & B & [  827.9,  83.0 ] &  10.8 &   0.5 &   1.9 &   0.4 & Hinkler\\
035-209-A & N1544832498 & B & [  608.8, 998.8 ] &  16.0 &   0.8 &   2.2 &   0.6 & Sikorsky\\
035-210-A & N1544832595 & B & [  617.1,  85.1 ] &  16.3 &   0.3 &   2.6 &   0.2 & Sikorsky\\
035-242-A & N1544835939 & B & [  836.7,  63.2 ] &  29.3 &   0.6 &   3.5 &   0.3 & Santos-Dumont\\
035-260-A & N1544837802 & B & [  699.5, 477.1 ] &  14.3 &   0.8 &   2.9 &   0.4 & Wright\\
035-296-A & N1544842159 & B & [  595.8, 129.0 ] &  26.0 &   1.6 &   3.5 &   0.4 & \\
035-299-A & N1544842586 & B & [ 1019.4, 405.8 ] &  50.6 &   1.3 &   5.4 &   0.2 & Bl\'eriot\\
035-299-A & N1544842586 & D &      "      & \multicolumn{2}{c}{} &   1.555 &  0.092 & Bl\'eriot\\
035-306-A & N1544843697 & B & [  612.5, 943.8 ] &  40.2 &   1.3 &   4.3 &   0.3 & Post\\
035-307-A & N1544843793 & B & [  601.4,  32.0 ] &  38.8 &   1.3 &   4.6 &   0.2 & Post\\
035-307-B &      "      & B & [  601.7, 161.1 ] &  43.9 &   1.2 &   6.0 &   0.2 & Earhart\\
\hline
036-070-A & N1546726798 & B & [  235.7,   3.7 ] &  88.8 &   2.0 &   3.7 &   0.6 & Bl\'eriot\\
\hline
037-001-A & N1548113882 & B & [  136.2, 732.2 ] & 145.7 &   5.2 &   4.9 &   1.0 & Bl\'eriot\\
037-002-A & N1548113946 & B & [  210.0, 665.4 ] & 122.9 &   5.0 &   7.0 &   0.9 & Bl\'eriot\\
\hline
039-009-A & N1551257093 & B & [  260.0,1001.1 ] & 112.2 &   3.7 &   5.0 &   1.0 & Bl\'eriot\\
039-140-A & N1551309107 & B & [  377.1, 994.7 ] & 104.9 &   3.2 &   5.7 &   0.9 & Bl\'eriot\\
\hline
041-088-A & N1552822117 & B & [  108.5, 912.0 ] & 112.0 &   5.8 &   3.1 &   1.8 & Bl\'eriot\\
041-089-A & N1552822477 & B & [  551.2, 998.7 ] & 123.2 &   4.0 &   4.1 &   1.3 & Bl\'eriot\\
\hline
042-045-A & N1554731052 & B & [  264.7, 484.6 ] &  16.3 &   1.8 &   2.1 &   0.4 & Richthofen\\
\hline
\multicolumn{9}{r}{Continued on next page}
\end{tabular}
\end{scriptsize}
\end{center}
\end{table*}

\setcounter{table}{3}
\begin{table*}[!h]
\begin{center}
\caption{Fitted sizes for \textit{trans}-Encke propellers (continued from previous page)}
\begin{scriptsize}
\begin{tabular}{ l c c c r @{$\pm$} l r @{$\pm$} l c }
\hline
\hline
Name$^a$ & Image & B/D$^b$ & [ line, sample ] & \multicolumn{2}{c}{$\Delta \ell$ (km)$^c$} & \multicolumn{2}{c}{$\Delta r$ (km)} & Match/Nickname$^d$\\
\hline
043-053-A & N1556183373 & B & [   74.7, 980.3 ] &  16.3 &   1.7 &   0.7 &   0.9 & 043-054-A\\
043-054-A & N1556183485 & B & [   80.4,  63.1 ] &  17.3 &   0.5 &   1.7 &   0.4 & 043-053-A\\
043-069-A & N1556185283 & B & [  390.0, 559.6 ] &  31.3 &   0.8 &   3.3 &   0.2 & \\
043-083-A & N1556186850 & B & [  343.4, 860.1 ] &  11.3 &   0.4 &   1.1 &   0.5 & \\
043-162-A & N1556196060 & B & [   64.8, 456.3 ] &  42.0 &   1.1 &   5.3 &   0.3 & Bl\'eriot\\
043-212-A & N1556202155 & B & [  333.1, 253.5 ] &  20.1 &   0.5 &   0.6 &   0.4 & \\
\hline
064-007-A & N1586618575 & B & [  999.3, 689.4 ] &   3.6 &   0.2 &   2.0 &   0.5 & \\
064-073-A & N1586624372 & B & [  332.5, 451.0 ] &  11.0 &   0.6 &   1.3 &   0.4 & \\
064-076-A & N1586624630 & B & [  330.4, 226.4 ] &   9.4 &   0.3 &   2.5 &   0.2 & \\
064-092-A & N1586626006 & B & [  315.0, 655.2 ] &   9.5 &   0.3 &   2.5 &   0.2 & \\
064-103-A & N1586627074 & B & [   35.6,  85.0 ] &  10.4 &   0.2 &   2.0 &   0.1 & 064-104-A\\
064-104-A & N1586627160 & B & [   72.0, 988.3 ] &  10.8 &   0.4 &   3.4 &   0.3 & 064-103-A\\
064-109-A & N1586627590 & B & [   21.8, 138.8 ] &  11.6 &   0.9 &   1.8 &   0.3 & Hinkler\\
064-114-A & N1586628020 & B & [   19.8, 675.4 ] &   9.7 &   0.3 &   2.3 &   0.2 & \\
064-121-A & N1586628622 & B & [  382.4, 550.5 ] &  37.6 &   0.8 &   4.9 &   0.1 & Earhart\\
064-133-A & N1586629654 & B & [  323.8, 501.9 ] &  18.6 &   1.0 &   2.5 &   0.2 & \\
064-146-A & N1586630893 & B & [  374.2, 264.5 ] &  19.4 &   0.2 &   3.1 &   0.1 & Post\\
064-163-A & N1586632355 & B & [  332.5, 182.9 ] &  11.5 &   0.6 &   3.4 &   0.2 & \\
064-176-A & N1586633595 & B & [  304.5, 464.1 ] &  15.0 &   0.3 &   1.3 &   0.1 & Lindbergh\\
064-208-A & N1586636591 & B & [  160.7, 359.5 ] &  25.7 &   0.4 &   5.3 &   0.2 & Kingsford Smith\\
064-222-A & N1586637920 & B & [  353.4, 231.8 ] &  10.3 &   0.3 &   2.3 &   0.2 & \\
064-254-A & N1586641169 & B & [   23.6,  59.9 ] &  48.5 &   0.7 &   5.5 &   0.1 & Bl\'eriot\\
064-255-A & N1586641255 & B & [   44.6, 971.5 ] &  50.5 &   0.7 &   5.5 &   0.1 & Bl\'eriot\\
\hline
068-017-A & N1589849522 & B & [  483.9, 329.5 ] &  54.7 &   2.6 &   5.5 &   0.9 & Bl\'eriot\\
068-017-A & N1589849522 & D &      "      & \multicolumn{2}{c}{} &   2.988 &  0.907 & Bl\'eriot\\
\hline
070-015-A & N1590907054 & D & [  249.8, 576.1 ] &  42.5 &   1.8 &   6.7 &   0.4 & Bl\'eriot\\
\hline
070-028-A & N1591067764 & B & [  501.4, 951.8 ] &  10.6 &   0.4 &   2.4 &   0.2 & \\
070-030-A & N1591068064 & B & [  658.8, 776.0 ] &  25.0 &   0.5 &   3.5 &   0.1 & \\
070-074-A & N1591074776 & B & [  357.1, 701.4 ] &  25.4 &   0.3 &   3.3 &   0.1 & \\
070-106-A & N1591080057 & B & [  713.3, 720.2 ] &  32.1 &   1.0 &   3.4 &   0.1 & \\
070-136-A & N1591085510 & B & [  558.7, 807.4 ] &  15.7 &   0.3 &   2.5 &   0.1 & \\
\hline
080-026-A & N1597462656 & B & [  850.6, 554.3 ] & 568.0 &  34.1 &   3.6 &   2.2 & Bl\'eriot\\
\hline
081-046-A & N1597788863 & D & [  550.6, 554.8 ] &   6.7 &   2.0 &   5.9 &   2.0 & Post\\
081-046-A & N1597788863 & B &      "      & \multicolumn{2}{c}{} &   2.970 &  0.428 & Post\\
081-054-A & N1597791119 & D & [  893.5, 162.6 ] &  18.2 &   2.2 &   6.2 &   0.5 & Bl\'eriot\\
081-054-A & N1597791119 & B &      "      & \multicolumn{2}{c}{} &   3.447 &  0.301 & Bl\'eriot\\
081-081-A & N1597800527 & D & [  580.6, 249.6 ] &   9.7 &   1.2 &   4.1 &   1.0 & Santos-Dumont\\
081-081-A & N1597800527 & B &      "      & \multicolumn{2}{c}{} &   2.920 &  0.325 & Santos-Dumont\\
\hline
090-017-A & N1603444112 & D & [   99.6, 556.2 ] &  22.8 &   1.6 &   6.9 &   0.5 & Bl\'eriot\\
090-017-A & N1603444112 & B &      "      & \multicolumn{2}{c}{} &   4.683 &  0.506 & Bl\'eriot\\
\hline
098-000-A & N1608800086 & B & [   84.2, 273.1 ] & 144.5 &   2.5 &   4.6 &   0.5 & Bl\'eriot\\
098-001-A & N1608800195 & B & [  192.7, 588.1 ] & 149.5 &   3.1 &   4.9 &   0.5 & Bl\'eriot\\
098-002-A & N1608800341 & B & [  333.1,1007.7 ] & 106.0 &   2.6 &   4.7 &   0.5 & Bl\'eriot\\
\hline
102-033-A & N1612496648 & B & [  625.7, 651.7 ] &  39.0 &   3.3 &   5.8 &   1.9 & Bl\'eriot\\
\hline
105-049-A & N1614861737 & B & [   77.5, 102.8 ] & \multicolumn{2}{c}{} &   4.123 &  1.215 & Bl\'eriot\\
\hline
116-003-A & N1628845563 & B & [  835.7, 168.2 ] &  28.0 &   3.0 &  10.3 &   1.3 & \\
116-004-A & N1628845780 & B & [  546.0, 620.0 ] &  22.1 &   2.9 &  10.4 &   1.2 & Santos-Dumont\\
116-005-A & N1628845813 & B & [  571.0, 590.7 ] &  29.6 &   3.0 &  11.6 &   1.2 & Santos-Dumont\\
116-008-A & N1628846480 & B & [  562.2, 678.3 ] &  31.4 &   1.4 &  11.8 &   0.7 & Earhart\\
116-009-A & N1628846513 & B & [  533.1, 727.2 ] &  31.9 &   1.4 &  12.0 &   0.6 & Earhart\\
\hline
\multicolumn{9}{l}{$^a$ Format is Orbit-Num-Letter, where Orbit and Num identify the image, and Letter identifies the feature}\\
\multicolumn{9}{l}{within the image \citep{Propellers08}.} \\
\multicolumn{9}{l}{$^b$ Relative-bright (B) or relative-dark (D), with respect to image background.}\\
\multicolumn{9}{l}{$^c$ If present, then fit was to a 2-D double-gaussian model \citep{Sremcevic07,Propellers08}.}\\
\multicolumn{9}{l}{If absent, then repeated 1-D gaussian fits were made in the radial direction \citep{Propellers06}.}\\
\multicolumn{9}{l}{$^d$ If the co-rotating location appears in more than one image, the corresponding locations are noted here.}\\
\end{tabular}
\end{scriptsize}
\end{center}
\end{table*}


\begin{thebibliography}{}

\bibitem[{Colwell} et~al., 2008]{ColwellEPSC08}
{Colwell}, J.~E., {Esposito}, L.~W., {Lissauer}, J.~J., {Jerousek}, R.~G., and
  {Srem{\v c}evi{\'c}}, M. (2008).
\newblock {Three-dimensional structure of Saturn's rings from Cassini UVIS
  stellar occultations}.
\newblock {\em European Planetary Science Congress Meeting Abstracts}, pages
  EPSC2008--A--00135.

\bibitem[{Colwell} et~al., 2009]{ColwellChapter09}
{Colwell}, J.~E., {Nicholson}, P.~D., {Tiscareno}, M.~S., {Murray}, C.~D.,
  {French}, R.~G., and {Marouf}, E.~A. (2009).
\newblock {The Structure of Saturn's Rings}.
\newblock In {Dougherty}, M., {Esposito}, L., and {Krimigis}, S.~M., editors,
  {\em Saturn from Cassini-Huygens}, pages 375--412. Springer-Verlag,
  Dordrecht.

\bibitem[{Hedman} et~al., 2007]{Hedman07}
{Hedman}, M.~M., {Nicholson}, P.~D., {Salo}, H., {Wallis}, B.~D., {Buratti},
  B.~J., {Baines}, K.~H., {Brown}, R.~H., and {Clark}, R.~N. (2007).
\newblock {Self-gravity wake structures in Saturn's A ring revealed by Cassini
  VIMS}.
\newblock {\em \aj}, 133:2624--2629.

\bibitem[{Lewis} and {Stewart}, 2009]{LS09}
{Lewis}, M.~C. and {Stewart}, G.~R. (2009).
\newblock {Features around embedded moonlets in Saturn's rings: The role of
  self-gravity and particle size distributions}.
\newblock {\em Icarus}, 199:387--412.

\bibitem[{Porco} et~al., 2007]{PorcoSci07}
{Porco}, C.~C., {Thomas}, P.~C., {Weiss}, J.~W., and {Richardson}, D.~C.
  (2007).
\newblock {Saturn's small satellites: Clues to their origins}.
\newblock {\em Science}, 318:1602--1607.

\bibitem[{Sei{\ss}} et~al., 2005]{Seiss05}
{Sei{\ss}}, M., {Spahn}, F., {Srem{\v c}evi{\'c}}, M., and {Salo}, H. (2005).
\newblock {Structures induced by small moonlets in Saturn's rings: Implications
  for the Cassini mission}.
\newblock {\em \grl}, 32:L11205.

\bibitem[{Spahn} and {Srem{\v c}evi{\'c}}, 2000]{SS00}
{Spahn}, F. and {Srem{\v c}evi{\'c}}, M. (2000).
\newblock {Density patterns induced by small moonlets in Saturn's rings?}
\newblock {\em \aap}, 358:368--372.

\bibitem[{Srem{\v c}evi{\'c}} et~al., 2007]{Sremcevic07}
{Srem{\v c}evi{\'c}}, M., {Schmidt}, J., {Salo}, H., {Sei{\ss}}, M., {Spahn},
  F., and {Albers}, N. (2007).
\newblock {A belt of moonlets in Saturn's A ring}.
\newblock {\em \nat}, 449:1019--1021.

\bibitem[{Srem{\v c}evi{\'c}} et~al., 2002]{SSD02}
{Srem{\v c}evi{\'c}}, M., {Spahn}, F., and {Duschl}, W.~J. (2002).
\newblock {Density structures in perturbed thin cold discs}.
\newblock {\em \mnras}, 337:1139--1152.

\bibitem[{Tiscareno} et~al., 2008]{Propellers08}
{Tiscareno}, M.~S., {Burns}, J.~A., {Hedman}, M.~M., and {Porco}, C.~C. (2008).
\newblock {The population of propellers in Saturn's A ring}.
\newblock {\em \aj}, 135:1083--1091.

\bibitem[{Tiscareno} et~al., 2006]{Propellers06}
{Tiscareno}, M.~S., {Burns}, J.~A., {Hedman}, M.~M., {Porco}, C.~C., {Weiss},
  J.~W., {Dones}, L., {Richardson}, D.~C., and {Murray}, C.~D. (2006).
\newblock {100-metre-diameter moonlets in Saturn's A Ring from observations of
  ``propeller'' structures}.
\newblock {\em \nat}, 440:648--650.

\bibitem[{Tiscareno} et~al., 2007]{soirings}
{Tiscareno}, M.~S., {Burns}, J.~A., {Nicholson}, P.~D., {Hedman}, M.~M., and
  {Porco}, C.~C. (2007).
\newblock {Cassini imaging of Saturn's rings II. A wavelet technique for
  analysis of density waves and other radial structure in the rings}.
\newblock {\em Icarus}, 189:14--34.

\bibitem[{Tiscareno} et~al., 2010]{Anparsgw10}
{Tiscareno}, M.~S., {Perrine}, R.~P., {Richardson}, D.~C., {Hedman}, M.~M.,
  {Weiss}, J.~W., {Porco}, C.~C., and {Burns}, J.~A. (2010).
\newblock {An analytic parameterization of self-gravity wakes in Saturn's
  rings}.
\newblock {\em \aj}, 139:492--503.

\bibitem[{Zebker} et~al., 1985]{Zebker85}
{Zebker}, H.~A., {Marouf}, E.~A., and {Tyler}, G.~L. (1985).
\newblock {Saturn's rings - Particle size distributions for thin layer model}.
\newblock {\em Icarus}, 64:531--548.

\end{thebibliography}
\end{document}